\documentclass[]{aa}
\usepackage[varg]{txfonts}
\usepackage{graphicx}
\usepackage{media9}
\usepackage{hyperref}
\usepackage{natbib}
\makeatletter
 \renewcommand\@biblabel[1]{}
\makeatother

\begin{document}
   \title{Crantor, a short-lived horseshoe companion to Uranus
          \thanks{Figures \ref{animation} and \ref{candidate} (animations) are 
                  available in electronic form at http://www.aanda.org}
         } 
   \author{C. de la Fuente Marcos
           \and
           R. de la Fuente Marcos}
   \authorrunning{C. de la Fuente Marcos \and R. de la Fuente Marcos}
   \titlerunning{Crantor, a horseshoe companion to Uranus 
                 }
   \offprints{C. de la Fuente Marcos, \email{nbplanet@fis.ucm.es}
              }
   \institute{Universidad Complutense de Madrid,
              Ciudad Universitaria, E-28040 Madrid, Spain}
   \date{Received 29 October 2012 / Accepted 2 January 2013}

   \abstract
      {Stable co-orbital motion with Uranus is vulnerable to planetary migration
       but temporary co-orbitals may exist today. So far only two candidates 
       have been suggested, both moving on horseshoe orbits: 83982~Crantor (2002 
       GO$_{9}$) and 2000~SN$_{331}$.  
       }
      {(83982)~Crantor is currently classified in the group of the Centaurs by 
       the MPC although the value of its orbital period is close to that of 
       Uranus. Here we revisit the topic of the possible 1:1 commensurability of 
       (83982)~Crantor with Uranus and also explore its dynamical past and look 
       into its medium-term stability and future orbital evolution. 
       }  
      {Our analysis is based on the results of $N$-body calculations that use 
       the most updated ephemerides and include perturbations by the eight major 
       planets, the Moon, the barycentre of the Pluto-Charon system, and the 
       three largest asteroids.
       }
      {(83982)~Crantor currently moves inside Uranus' co-orbital region on a 
       complex horseshoe orbit. The motion of this object is primarily driven by 
       the influence of the Sun and Uranus, although Saturn plays a significant 
       role in destabilizing its orbit. The precession of the nodes of (83982) 
       Crantor, which is accelerated by Saturn, controls its evolution and 
       short-term stability. Although this object follows a temporary horseshoe 
       orbit, more stable trajectories are possible and we present 2010 
       EU$_{65}$ as a long-term horseshoe librator candidate in urgent need of 
       follow-up observations. Available data indicate that the candidate 2000 
       SN$_{331}$ is not a Uranus' co-orbital.
       }
      {Our calculations confirm that (83982)~Crantor is currently trapped in the
       1:1 commensurability with Uranus but it is unlikely to be a primordial 
       1:1 librator. Although this object follows a chaotic, short-lived 
       horseshoe orbit, longer term horseshoe stability appears to be possible.  
       We also confirm that high order resonances with Saturn play a major role 
       in destabilizing the orbits of Uranus co-orbitals.
       }


   \maketitle

   \section{Introduction}
      An object librating in a trajectory which encompasses the Lagrangian points L$_4$, L$_5$ and L$_3$ of a host planet evolves 
      on a so-called regular horseshoe orbit. Viewed in a frame of reference that co-rotates with the host planet, the shape of 
      such an orbit projected onto the ecliptic plane resembles that of an actual horseshoe although its three-dimensional layout 
      looks more like a corkscrew around the orbit of the host planet while both revolve around the Sun. The size of the horseshoe
      orbit depends on the mass of the host planet, being wider and having a better chance of survival when this mass decreases 
      (Dermott \& Murray 1981a). In general, horseshoe orbits are not considered to be long-term stable (Dermott \& Murray 1981a; 
      Murray \& Dermott 1999). Horseshoe orbits were originally predicted by Brown (1911) and Darwin (1912) and further studied 
      later by, for example, Thuring (1959), Rabe (1961), Giacaglia (1970), Weissman \& Wetherill (1974) and Garfinkel (1977) but 
      were largely considered theoretical curiosities until the Saturnian moons Janus and Epimetheus were identified as horseshoe 
      librators (Smith et al. 1980; Synnott et al. 1981; Dermott \& Murray 1981b). The existence of minor planets moving on 
      horseshoe orbits around the major planets was first postulated by Milani et al. (1989) and Michel et al. (1996) on 
      theoretical grounds. 

      In the solar system, there are several real examples of minor bodies moving on such orbits. The first minor body to be 
      confirmed to follow a horseshoe orbit was 3753 Cruithne (1986 TO) (Wiegert et al. 1997, 1998), in this case with the Earth. 
      Karlsson (2004) found multiple objects moving in temporary horseshoe orbits with Jupiter. Connors et al. (2004) identified 
      an object, 2003 YN$_{107}$, following a compound horseshoe-quasi-satellite orbit with the Earth. Additional objects moving 
      in comparable trajectories are 2002 AA$_{29}$ and 2001 GO$_{2}$ (Brasser et al. 2004). 2001 CK$_{32}$ follows an orbit 
      similar to that of (3753) Cruithne but hosted by Venus, not the Earth (Brasser et al. 2004). (36017) 1999 ND$_{43}$ is a 
      horseshoe librator with Mars (Connors et al. 2005). Yet another horseshoe companion of the Earth was found in 2010 SO$_{16}$ 
      (Christou \& Asher 2011). Additional horseshoe librators with Jupiter were recently identified by Wajer \& Kr\'olikowska 
      (2012). Finally, (310071) 2010 KR$_{59}$ is following a temporary and rather complex horseshoe orbit with Neptune (de la 
      Fuente Marcos \& de la Fuente Marcos 2012a).

      Numerical simulations predict that Uranus may have retained a certain amount of its primordial co-orbital minor planet
      population (Holman \& Wisdom 1993; Wiegert et al. 2000; Nesvorn\'y \& Dones 2002; Marzari et al. 2003). However, Uranus 
      appears not to be able to efficiently capture objects into the 1:1 commensurability today even for short periods of time 
      (Horner \& Evans 2006). The stability of hypothetical Uranus co-orbitals, specifically those moving in tadpole orbits,
      has been studied by Dvorak et al. (2010) and they have found that the orbital inclination is the key parameter regarding
      stability, only the inclination intervals (0, 7)$^{\circ}$, (9, 13)$^{\circ}$, (31, 36)$^{\circ}$ and (38, 50)$^{\circ}$  
      appear to be stable. This scarcity of Uranus co-orbitals seems to be confirmed by current observational results. Although 
      hundreds of objects have been discovered in the outer solar system during the various wide-field surveys carried out during 
      the past decade, the two objects pointed out by Gallardo (2006) remain as the only Uranus' co-orbital candidates identified 
      to date.   

      Calculations by Gallardo (2006) revealed that two objects were moving in a 1:1 mean motion resonance with Uranus. One of 
      them, 83982 Crantor (2002 GO$_{9}$), is the main object of study of this paper. Since 2006, the orbit of this object has 
      been improved and here we make use of the most updated ephemerides to reassess the current dynamical status of this minor 
      body. In this paper, we use $N$-body simulations to confirm the co-orbital nature with Uranus of the asteroid (83982) 
      Crantor, currently classified as Centaur by both the Minor Planet Center (MPC) and the Jet Propulsion Laboratory (JPL). The 
      numerical model is described in the next section and available data on (83982) Crantor are presented in Section 3. The 
      results of our $N$-body calculations are shown in Section 4. These results are discussed in Section 5. A new Uranus' 
      horseshoe librator candidate is presented in Section 6 and our conclusions are summarized in Section 7.

   \section{Numerical integration}
      The orbital evolution of 83982 Crantor (2002 GO$_{9}$) was computed for 0.5 Myr forward and backward in time using the 
      Hermite integration scheme described by Makino (1991) and implemented by Aarseth (2003). This $N$-body code has been 
      extensively tested by the authors and used in a variety of recent solar system numerical studies (de la Fuente Marcos \& de 
      la Fuente Marcos 2012a,b,c,d). The standard version of this sequential code is publicly available from the IoA web 
      site\footnote{\url{http://www.ast.cam.ac.uk/~sverre/web/pages/nbody.htm}}. Our integrations include the perturbations by the 
      eight major planets, the Moon, the barycentre of the Pluto-Charon system, and the three largest asteroids. For accurate 
      initial positions and velocities we used the heliocentric ecliptic Keplerian elements provided by the Jet Propulsion 
      Laboratory on-line solar system data service\footnote{\url{http://ssd.jpl.nasa.gov/?planet\_pos}} (Giorgini et al. 1996) and 
      initial positions and velocities based on the DE405 planetary orbital ephemerides (Standish 1998) referred to the barycentre 
      of the solar system. In addition to the orbital calculations completed using the nominal elements in Table \ref{elements}, 
      we have performed 50 control simulations with sets of orbital elements obtained from the nominal ones and the quoted 
      uncertainties (3-$\sigma$). The derived sample of control orbits follows a Gaussian distribution in the 6-dimensional space 
      of orbital elements and they are compatible with the observations within the 3-$\sigma$ uncertainties. The analysis of the 
      control orbits provides some insight on the predictability of the trajectory of the object. The numerical rather than 
      analytical approach to the study of this object is more appropriate because planets other than Uranus, as we will discuss 
      later, play a role on the dynamics of this asteroid, rendering the traditional perturbational approach within the framework 
      of the three-body problem particularly limited in this case. There exists practically no analytical means to study horseshoe
      orbits.

%
%
         \begin{table}
          \fontsize{8}{11pt}\selectfont
          \tabcolsep 0.35truecm
          \caption{Heliocentric Keplerian orbital elements of 83982 Crantor (2002 GO$_{9}$) used in this research. Values include 
                   the 1-$\sigma$ uncertainty. The orbit is based on 104 observations spanning a data-arc of 2,654 days or 7.27 yr,
                   from 2001-03-20 to 2008-06-25.
                   (Epoch = JD2456200.5, 2012-Sep-30.0; J2000.0 ecliptic and equinox. Source: JPL Small-Body Database.)
                  }
          \begin{tabular}{ccc}
           \hline
            semi-major axis, $a$                        & = & 19.3553$\pm$0.0014 AU \\
            eccentricity, $e$                           & = & 0.27496$\pm$0.00004 \\
            inclination, $i$                            & = & 12.78489$\pm$0.00003 $^{\circ}$ \\
            longitude of the ascending node, $\Omega$   & = & 117.4097$\pm$0.0003 $^{\circ}$ \\
            argument of perihelion, $\omega$            & = & 92.599$\pm$0.003 $^{\circ}$ \\
            mean anomaly, $M$                           & = & 43.964$\pm$0.006 $^{\circ}$ \\
            perihelion, $q$                             & = & 14.0333$\pm$0.0002 AU \\
            aphelion, $Q$                               & = & 24.677$\pm$0.002 AU \\
            absolute magnitude, $H$                     & = & 8.5$\pm$0.8 mag \\
           \hline
          \end{tabular}
          \label{elements}
         \end{table}
%
%

   \section{83982 Crantor (2002 GO$_{9}$) in perspective}
      83982 Crantor (2002 GO$_{9}$) was discovered on April 12, 2002 by E.~F. Helin, S. Pravdo, K. Lawrence, M. Hicks and R. 
      Thicksten working for the Near-Earth Asteroid Tracking (NEAT) project at Palomar Observatory (Gilmore et al. 2002). It was
      originally reported as a scattered disk object with $a$ = 54.24 AU and $e$ = 0.81 but soon after, a number of precovery 
      images of the object were uncovered: it first appears in images obtained on March 20, 2001 from the Air Force Maui Optical 
      and Supercomputing (AMOS) observatory located at the summit of Haleakala, then on images acquired on April 16, 2001 from the 
      Apache Point Observatory as part of the Sloan Digital Sky Survey (SDSS), and again on new images obtained on March 26, 2001 
      and January 15, 2002 from Haleakala-AMOS (Ticha et al. 2002). All this observational material enabled the computation of a 
      reliable orbit characterized by a value of the semi-major axis (19.36 AU) close to that of Uranus, significant eccentricity 
      ($\sim$0.3), and moderate inclination ($\sim$13$^{\circ}$). Therefore, its orbit is now relatively well determined with 104 
      observations spanning a data-arc of 2,654 days and it is clearly not compatible with that of a scattered disk object. 
      Consistently, (83982) Crantor is currently listed by both the Minor Planet Center (MPC) 
      Database\footnote{\url{http://www.minorplanetcenter.net/db\_search}} and the JPL Small-Body 
      Database\footnote{\url{http://ssd.jpl.nasa.gov/sbdb.cgi}} as a Centaur. 

      Early photometric work (Tegler et al. 2003) pointed out the red surface color of (83982) Crantor and suggested that this
      object and many others like it were formed farther away from the Sun than distant minor bodies characterized by grey surface 
      color. In fact, (83982) Crantor is one of the reddest objects of the solar system, close to fellow ultra-red Centaur (5145) 
      Pholus; its surface should be partially covered by tholins in order to explain its redness and low albedo (Cruikshank et al. 
      2007). The rotation period of (83982) Crantor is 6.97 h or 9.67 h with a light-curve amplitude of 0.14 mag (Ortiz et al. 
      2003). (83982) Crantor was observed with the near-infrared integral field spectrograph SINFONI at the Very Large Telescope 
      (VLT) that found evidence of an absorption feature in its spectrum at 2.0 $\mu$m, probably associated with water ice and 
      another feature at 2.3 $\mu$m which could be associated with methanol (Alvarez-Candal et al. 2007). These results confirmed 
      previous hints obtained by Doressoundiram et al. (2005). Visible spectra further support the very red nature of the object 
      (Alvarez-Candal et al. 2008). Additional near-infrared spectra taken with the Keck I Telescope confirmed previous results 
      (Barkume et al. 2008). Near-infrared photometry was obtained by Doressoundiram et al. (2007). Incomplete photometry was 
      obtained with FORS1 at VLT (DeMeo et al. 2009). Recent Hubble Wide Field Camera 3 results (Fraser \& Brown 2012) validate 
      previous findings, indicating the presence of irradiated organics and tholins on its surface.

      The asteroid (83982) Crantor is relatively large. The object has a diameter of $<66.7^{+18.7}_{-19.6}$ km with visible 
      geometric albedo of 8.60$^{+8.62}_{-3.36}$\% (Stansberry et al. 2008). Its period of revolution around the Sun, 
      approximately 85.15 yr at present, is very close to that of Uranus, 84.32 yr. As a result, (83982) Crantor and Uranus appear 
      to follow each other in their paths around the Sun, although (83982) Crantor's orbital plane is currently tilted to that of 
      the Earth by 12.8$^{\circ}$ (Uranus' is 0.8$^{\circ}$). Its dynamical half-lives have been estimated to be 2.93 Myr (for 
      forward integration) and 3.67 Myr (for backward integration) by Horner et al. (2004). (83982) Crantor was originally 
      proposed as a possible co-orbital of Uranus together with 2000 SN$_{331}$ by Gallardo (2006). Both objects would be 
      following horseshoe trajectories. In the same research work, it is pointed out that Uranus' Trojans are affected by high 
      order resonances with Saturn. The asteroid 2000 SN$_{331}$ has not been reobserved since its discovery and its orbit remains 
      very poorly known (see below). In the following section we focus on the dynamical evolution of (83982) Crantor.

   \section{Dynamical evolution}
      In order to study the librational properties of 83982 Crantor (2002 GO$_{9}$) and following the work of Mikkola et al. 
      (2006), we define the relative deviation of the semi-major axis from that of Uranus by $\alpha = (a - a_U) / a_U$, where $a$ 
      and $a_U$ are the semi-major axes of the object and Uranus, respectively, and also the relative mean longitude $\lambda_r = 
      \lambda - \lambda_U$, where $\lambda$ and $\lambda_U$ are the mean longitudes of the object and Uranus, respectively. If 
      $\lambda_r$ oscillates around 0$^{\circ}$, the object is considered a quasi-satellite; Trojan bodies are characterized by 
      $\lambda_r$ oscillating around +60$^{\circ}$ (L$_4$ Trojan) or -60$^{\circ}$ (or 300$^{\circ}$, L$_5$ Trojan); finally, 
      an object librating with amplitude $> 180^{\circ}$ follows a horseshoe orbit (see, e.g., Murray \& Dermott 1999). 

\begin{figure}
       \centering
        \includegraphics[width=\linewidth]{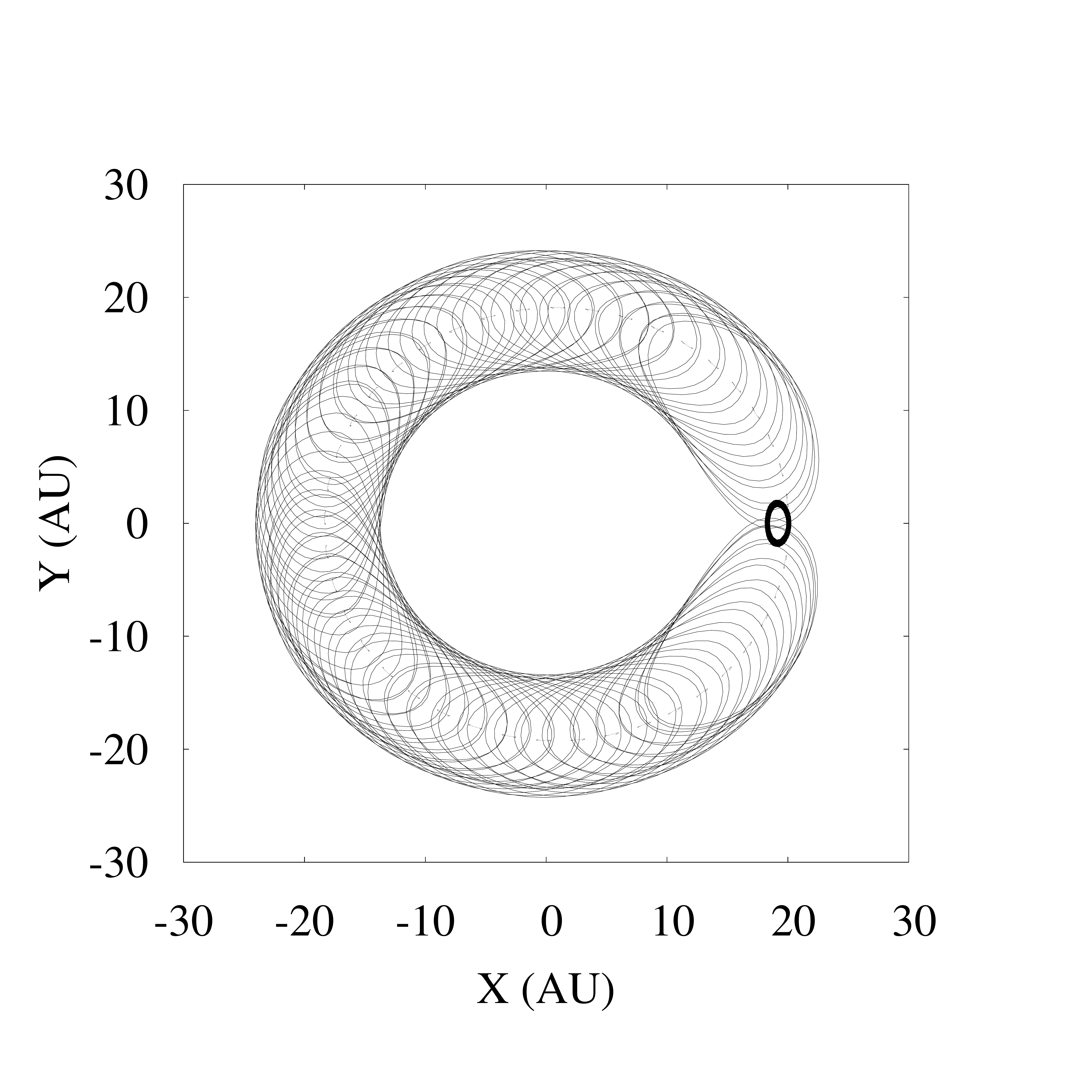}
        \caption{The motion of 83982 Crantor (2002 GO$_{9}$) over the time range (-3, 5) kyr is displayed projected onto the 
                 ecliptic plane in a coordinate system rotating with Uranus. The orbit and the position of Uranus are also 
                 indicated. In this frame of reference and as a result of its non-negligible eccentricity, Uranus describes a 
                 small ellipse. The trajectory of the object spiralizes along Uranus' orbit at a rate of nearly 0$\fdg$08 per 
                 year.
                }
        \label{hs}
     \end{figure}

\begin{figure*}
        \centering
        \includemedia[
          label=Crantor,
          width=\textwidth,height=0.65\textwidth,
          activate=onclick,
          addresource=Crantor3DBw.mp4,
          flashvars={
            source=Crantor3DBw.mp4
           &autoPlay=true
           &loop=true
           &controlBarMode=floating
           &controlBarAutoHide=false
           &scaleMode=letterbox
          }
        ]{\includegraphics{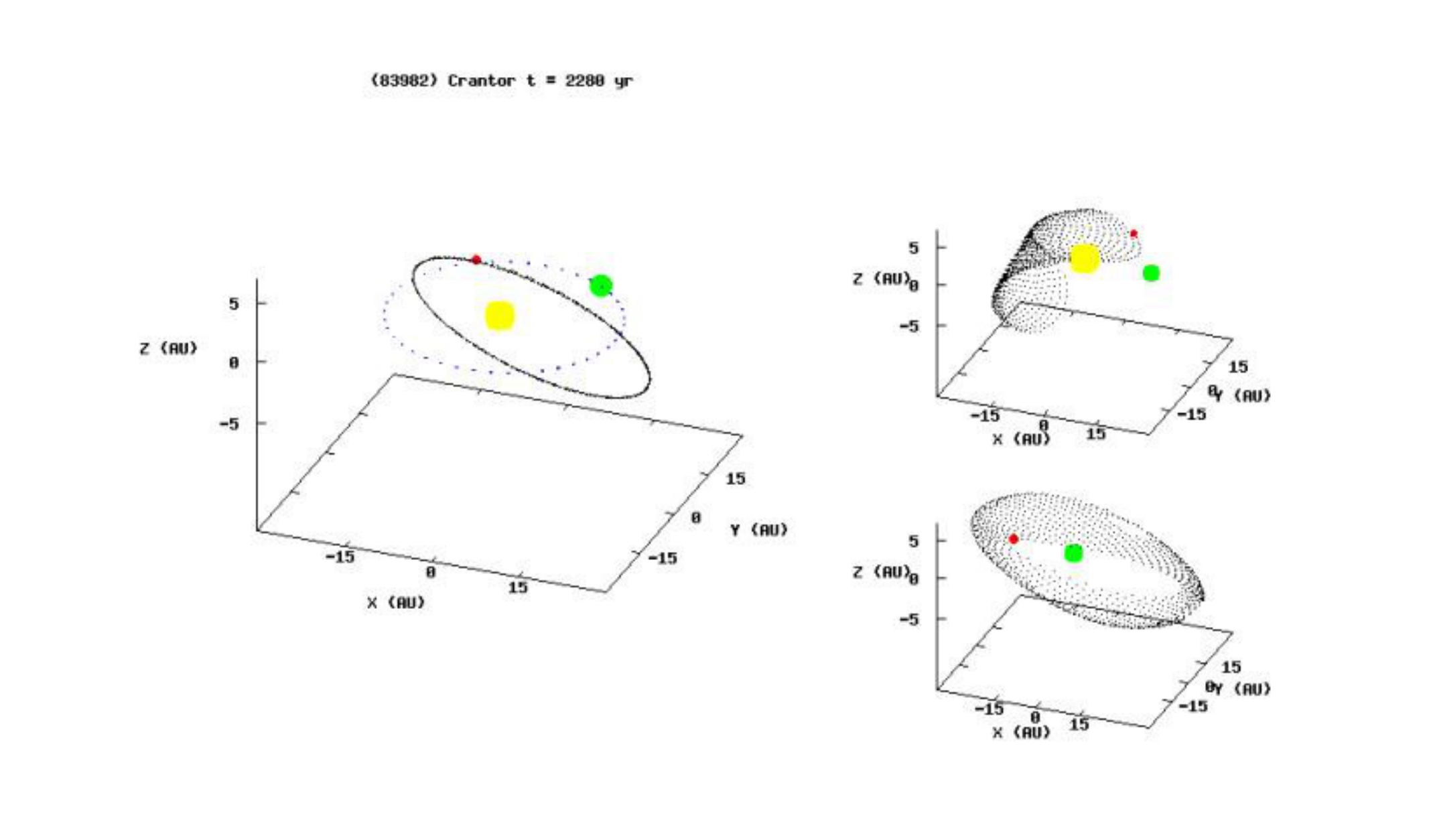}}{VPlayer.swf}
        \PushButton[
           onclick={
             annotRM['Crantor'].activated=true;
             annotRM['Crantor'].callAS('play');
           }
        ]{\fbox{Play}}
        \PushButton[
           onclick={
             annotRM['Crantor'].activated=true;
             annotRM['Crantor'].callAS('pause');
           }
        ]{\fbox{Pause}}
        \caption{Three-dimensional evolution of the orbit of 83982 Crantor (2002 GO$_{9}$) in three different frames of reference: 
                 heliocentric (left), frame co-rotating with Uranus but centered on the Sun (top-right), and Uranocentric 
                 (bottom-right). The red point is (83982) Crantor, the green one is Uranus, and the yellow one is the Sun. The
                 osculating orbits are outlined and the viewing angle changes slowly to make easier for the reader to visualize
                 the orbital evolution.} 
        \label{animation}
     \end{figure*}

      Our $N$-body calculations confirm that (83982) Crantor currently is a co-orbital companion to Uranus and follows a horseshoe 
      orbit, all in agreement with what was originally pointed out by Gallardo (2006), see Fig. \ref{hs}. The apparent overlap 
      with Uranus' position in Fig. \ref{hs} is the result of the moderate orbital inclination of the object. The orbital 
      behaviour of (83982) Crantor is illustrated by the animation displayed in Fig. \ref{animation} (available on the electronic 
      edition as a high resolution animation or embedded at lower resolution in the pdf file associated to this paper). The orbit 
      is presented in three frames of reference: heliocentric (left), co-rotating with Uranus (top-right) and Uranocentric 
      (bottom-right). (83982) Crantor moves in a non-regular horseshoe orbit with a period of about 8500 years. Non-regular means 
      that compound horseshoe-quasi-satellite loops are possible. The dynamical evolution of an object moving in a horseshoe orbit 
      associated to Uranus can be decomposed into a slow guiding centre motion and a superimposed short period three-dimensional 
      epicyclic motion viewed in a frame of reference co-rotating with Uranus. The object spirals along Uranus' orbit at a rate of 
      nearly 0$\fdg$08 per year, but each time it gets close to Uranus is effectively repelled by the planet. The reversals of the 
      net motion of the object with respect to Uranus are obvious in the accompanying animation and given the fact that they take 
      place when the distance to Uranus is the smallest, the gravitational interaction with Uranus is, at that moment, the 
      strongest. Although the inclination of the asteroid is high enough to avoid close encounters with Uranus when the relative 
      mean longitude approaches zero, these close encounters that can only occur in the vicinity of the nodes play a major role on 
      the activation and deactivation of the horseshoe behaviour of this object (see below). 
%
%
%
%
%
%
%
%

\begin{figure}
       \centering
        \includegraphics[width=\linewidth]{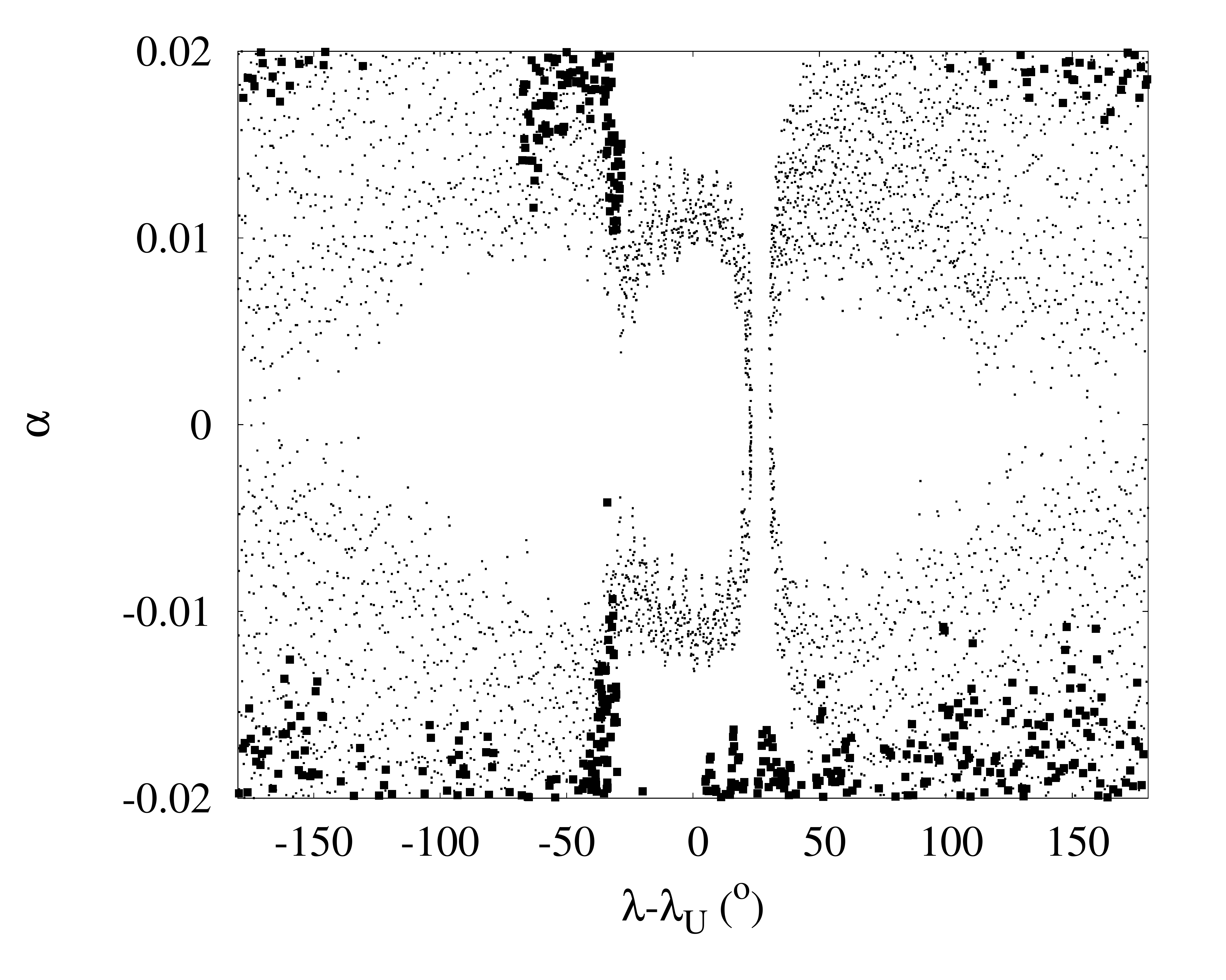}
        \caption{Resonant evolution of the asteroid 83982 Crantor (2002 GO$_{9}$). The relative deviation of its semi-major axis 
                 from that of Uranus, $\alpha$, as a function of the relative mean longitude, $\lambda_r$, during the time 
                 intervals (-22, -10) kyr (black squares) and (2, 12) kyr (dots) is displayed. The first interval covers the close 
                 encounter observed in Fig. \ref{all}, panel A (see the text). The second interval lasts an entire cycle of the 
                 horseshoe orbit and includes a quasi-satellite half libration, i.e., $\lambda_r$ goes beyond -60$^{\circ}$, 
                 passes 0$^{\circ}$ but does not reach 60$^{\circ}$.                 
                }
        \label{f2}
     \end{figure}

      The relative deviation of the semi-major axis, $\alpha$, as a function of the relative mean longitude, $\lambda_r$, is 
      displayed in Fig. \ref{f2}, for selected time intervals. The relative mean longitude of (83982) Crantor librates around the 
      unstable Lagrangian point L$_3$ at 180$^{\circ}$ with large amplitude, allowing the object to come quite close to Uranus and 
      to suffer destabilizing close encounters. Typical horseshoe behaviour is observed but brief (half a loop) quasi-satellite 
      episodes also take place. During these events, the object still moves in a 1:1 commensurability with Uranus but $\lambda_r$ 
      librates about 0$^{\circ}$. This exchange between horseshoe and quasi-satellite paths (or compound horseshoe-quasi-satellite
      orbits) is observed in other horseshoe librators, for example 3753 Cruithne (1986 TO) (Wiegert et al. 1997, 1998). 

%
%
%
%

\begin{figure}
       \centering
        \includegraphics[width=\linewidth]{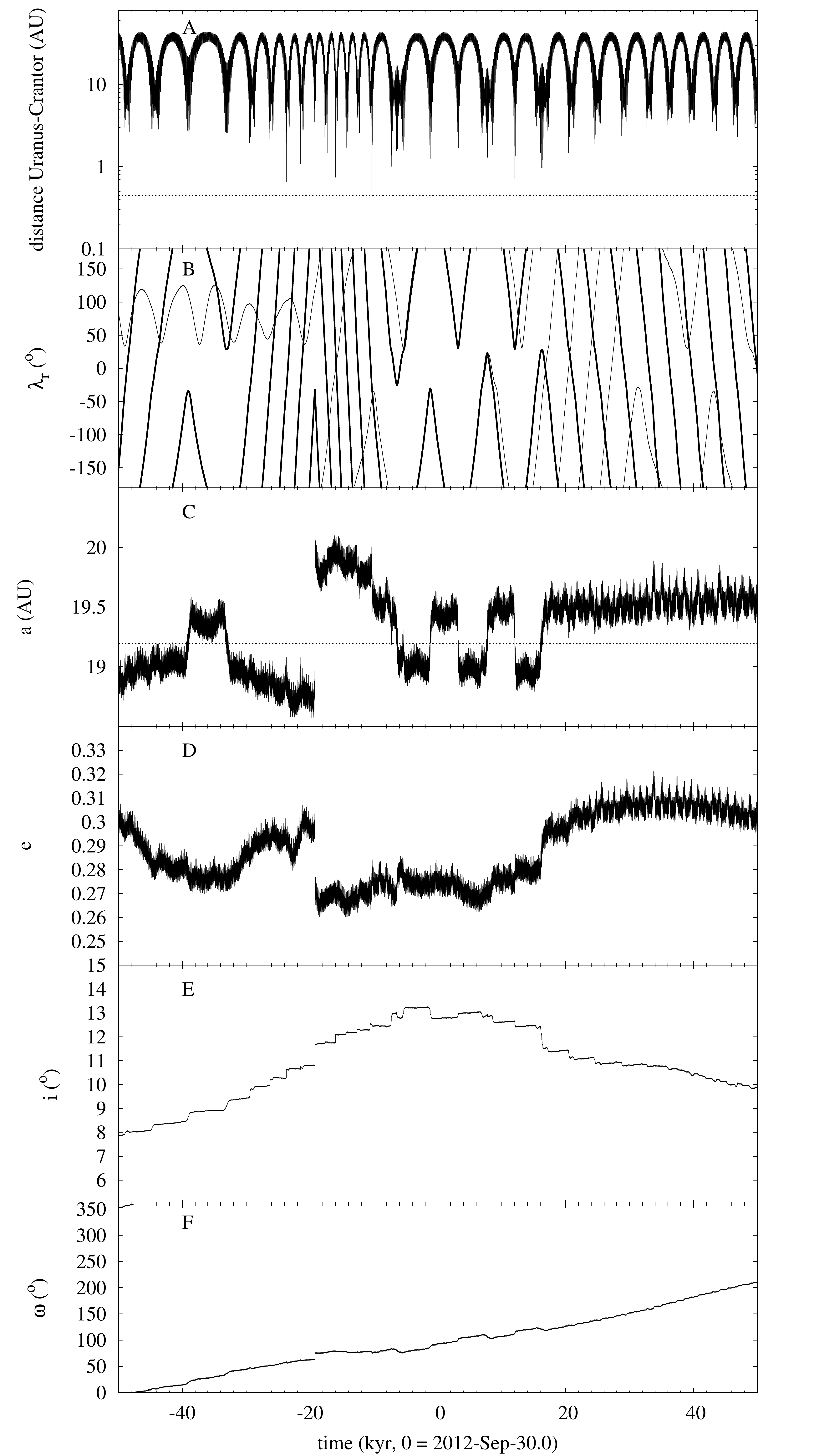}
        \caption{Time evolution of various parameters. The distance of 83982 Crantor (2002 GO$_{9}$) from Uranus (panel A); the
                 value of the Hill sphere radius of Uranus, 0.447 AU, is displayed. The resonant angle, $\lambda_{r}$ (panel B)
                 for the nominal orbit in Table \ref{elements} (thick line) and one of the control orbits (thin line). This
                 particular control orbit has been chosen close to the 3-$\sigma$ limit so its orbital elements are most different
                 from the nominal ones. The orbital elements $a$ (panel C) with the current value of Uranus' semi-major axis,
                 $e$ (panel D), $i$ (panel E), and $\omega$ (panel F). 
                }
        \label{all}
     \end{figure}

      A plot of the orbital elements of (83982) Crantor over a 100 kyr interval centred on the present is shown in Fig. \ref{all}. 
      The distance of (83982) Crantor from Uranus displayed in Fig. \ref{all}, panel A shows that the object undergoes close
      encounters with Uranus. A very close encounter, almost a collision at 0.17 AU or 0.37 Hill radii (which is 0.45 AU for 
      Uranus), took place 19248 yr ago. The evolution of $\lambda_{r}$ in panel B indicates that the current horseshoe episode
      will end in about 20000 yr from now; (83982) Crantor will decouple from Uranus with $\lambda_{r}$ circulating not librating.  
      The semi-major axis exhibits an oscillatory behaviour that is characteristic of the effects of a 1:1 mean motion resonance. 
      The eccentricity decreases by 10\% during the very close encounter pointed out above and then increases again after nearly 
      19000 yr, see Fig. \ref{all}, panel D. During the horseshoe episode, the orbital inclination remains in the interval (12, 
      13)$^{\circ}$, see Fig. \ref{all}, panel E. The argument of perihelion, Fig. \ref{all}, panel F, circulates. In Fig. 
      \ref{all}, panel B, we also show the evolution of the relative mean longitude for a particular control orbit that has been 
      chosen close to the 3-$\sigma$ limit; in this way, its orbital elements are most different from the nominal ones in Table 
      \ref{elements}. For this particular orbit and prior to entering the horseshoe dynamical state, the object was an L$_4$ 
      Trojan. All the control calculations indicate that (83982) Crantor has been co-orbital with Uranus for at least 100 kyr.
        
%
%
%
%

   \section{Discussion}
      The characteristic e-folding time of 83982 Crantor (2002 GO$_{9}$) during the present horseshoe dynamical state has been 
      found to be of order of 1 kyr. This value gives the reader the idea of how small the timescale required for two initially
      infinitesimally close trajectories to separate significantly is in this case. An additional test, that confirms how 
      sensitive to small changes the dynamical evolution of (83982) Crantor is, can be obtained by repeating the calculations, 
      this time excluding one of the three largest asteroids; the orbit significantly diverges from the standard one after just a 
      few 10 kyr. Therefore, simulations over long timescales (e.g. 1 Myr) are not appropriate in this case and because of that, 
      we restrict our figures to a few 10 kyr. Since the orbit of the asteroid is chaotic, its true phase-space trajectory will 
      diverge exponentially from that obtained in our calculations. However, the evolution of the control orbits exhibits very 
      similar secular behaviour of the orbital elements in the time interval (-10, 10) kyr. We also note that at the ends of the 
      interval, a close encounter between the asteroid and Uranus happened. Therefore, the dynamical evolution of (83982) Crantor 
      as described by our integrations can be considered reliable within that relatively short time interval but outside, we 
      should regard our results as an indication of the probable dynamical behaviour of the object. 

      Our calculations show that the orbit of (83982) Crantor is predictable only within a relatively short time interval. We 
      consider that an orbit is predictable when all the control orbits give comparable results. It is clear that the current 
      horseshoe behaviour is not stable on a timescale longer than a few 10 kyr. The analysis of the control orbits indicates that 
      this object may have been co-orbital for less than 200 kyr and it may leave Uranus' co-orbital region in less than 100 kyr. 
      However, some control orbits remain in the co-orbital region for about 1 Myr. About half the studied control orbits started 
      co-orbital motion about 10-20 kyr ago but one third left the circulation regime to become co-orbitals about 80 kyr ago. The 
      rest appear to have been switching between the various co-orbital states for several 100 kyr. The changes in dynamical state 
      are always associated to close encounters with Uranus. Regarding the future orbital evolution of (83982) Crantor, the 
      majority (70\%) of control orbits continue in the horseshoe state for about 15 kyr. The remaining orbits divide evenly among 
      those lasting less and those lasting more than 15 kyr. Twenty per cent of control orbits leave the co-orbital region 
      permanently after ending their horseshoe stage but nearly 70\% return after 5 to 15 kyr. The rest switch to the 
      quasi-satellite phase. Out of the returning co-orbitals, the vast majority make a comeback as horseshoe librators with only 
      15\% becoming quasi-satellites or Trojans. Nearly 25\% continue switching between the various co-orbital states for several 
      100 kyr and the rest leave the co-orbital region after nearly 80 kyr. Among the co-orbital states, the vast majority of the 
      episodes produce irregular horseshoe orbits and they also last longer; quasi-satellite events are a distant second and 
      tadpole orbits are observed in just a few cases. Most discrete episodes last for less than 10 kyr. As the reader can see, it 
      is rather difficult to provide a clean picture of the medium-term past and future of this object. The amount of time spent 
      in the horseshoe state prior to the origin of time considered in this research is similar in all the control models and in 
      most of them the relative mean longitude circulates before entering and after leaving the state but the details beyond 20 
      kyr into the past and after 30 kyr into the future are quite heterogeneous both in terms of the actual type of orbital 
      behaviour and its duration. 

\begin{figure}
       \centering
        \includegraphics[width=\linewidth]{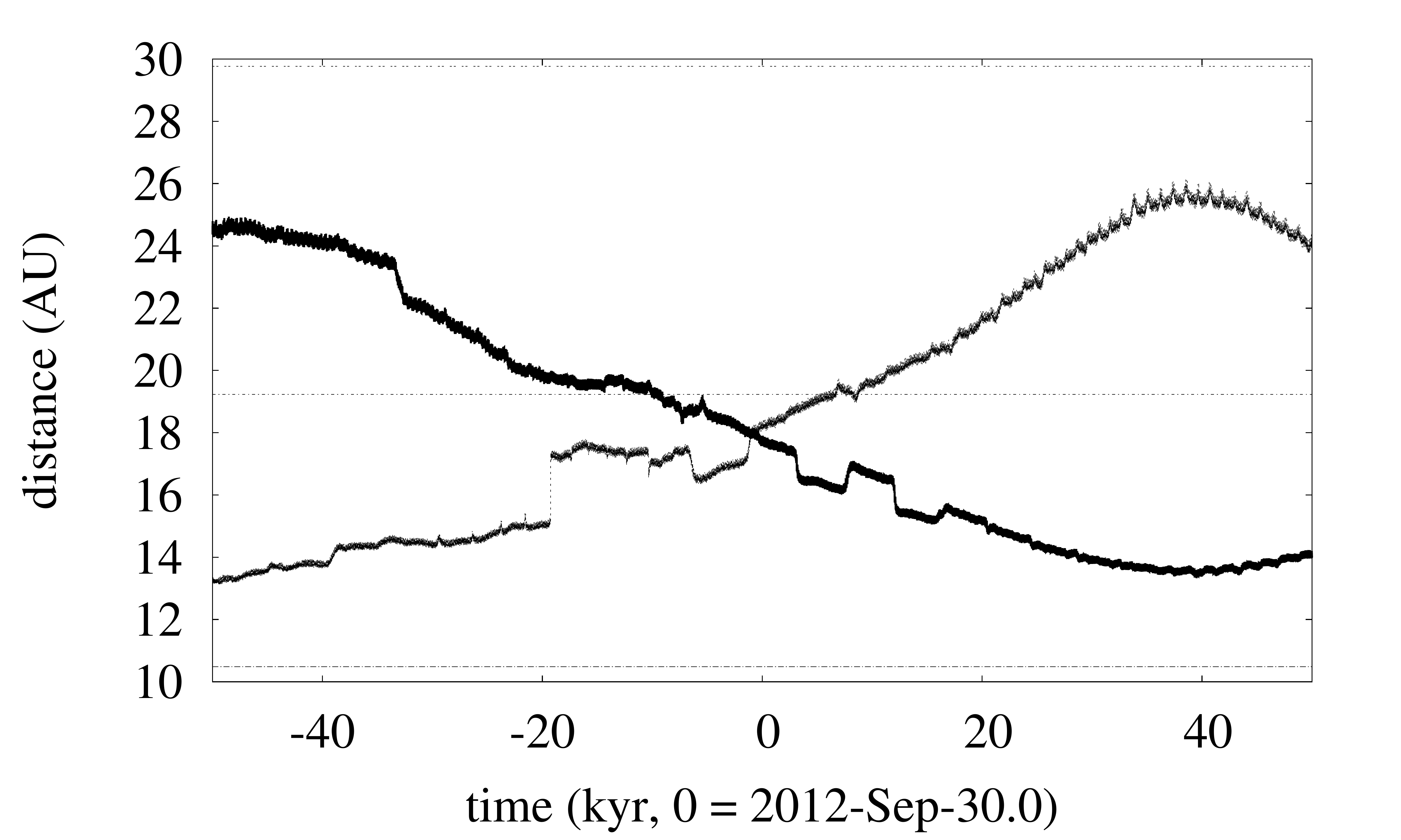}
        \caption{Heliocentric distance to the descending (thick line) and ascending nodes (dotted line) of 83982 Crantor (2002 
                 GO$_{9}$). Saturn's aphelion, Uranus' semi-major axis, and Neptune's perihelion distances are also shown. Both 
                 nodal distances are at present relatively close to the value of Uranus' semi-major axis which explains why close 
                 encounters with Uranus are possible and relatively frequent, but most of the time the asteroid remains at a safe 
                 distance from Uranus.
                }
        \label{node}
     \end{figure}

      The asteroid 83982 Crantor (2002 GO$_{9}$) follows an eccentric orbit ($e\approx0.3$) but it currently crosses only the 
      orbit of Uranus. In general, minor bodies that cross the paths of one or more planets can be rapidly destabilized by 
      scattering resulting from close planetary approaches if their orbital inclinations are small. (83982) Crantor moves in a 
      relatively highly inclined orbit ($i \approx 13^{\circ}$). In the solar system and for a minor body moving in an inclined 
      orbit, close encounters with major planets are only possible in the vicinity of the nodes. The distance between the Sun and 
      the nodes is given by $r = a (1 - e^2) / (1 \pm e \cos \omega)$, where the "+" sign is for the ascending node and the "-" 
      sign is for the descending node. Figure \ref{node} shows the evolution of the distance to the nodes of (83982) Crantor in 
      the time range (-50, 50) kyr. The evolution of the orbital elements in Fig. \ref{all} shows that changes in $\omega$
      dominate those in $a$ and $e$ for (83982) Crantor and largely control the positions of the nodes. The current precession
      rate of the nodes is nearly +0$\fdg$2 per century. This value decreases when $\lambda_r$ circulates. We found that, in 
      accordance with theory (Namouni 1999), if (83982) Crantor moves in a horseshoe orbit, $\dot{\omega} > 0$ but during the 
      brief quasi-satellite loops, the value of the argument of perihelion decreases (see Fig. \ref{all}, panel F). The values of 
      the nodal distances are currently very close to the value of the semi-major axis of Uranus. Close encounters are therefore 
      possible at both nodes doubling the probability of having a strong gravitational interaction with Uranus that may 
      significantly change the orbit. When the object is east of Uranus, encounters occur at the ascending node. In contrast, 
      close encounters take place at the descending node when (83982) Crantor approaches Uranus from the west. In this way, the 
      gravitational perturbations from Uranus are most effective and both Saturn and Neptune are secondary perturbers for this 
      object. Currently and as a result of its horseshoe trajectory, the object approaches Uranus every 4354 yr although most of 
      the time the asteroid remains at a safe distance from Uranus.
%
%
%
%

      Although (83982) Crantor's eccentricity is not large enough to cross Saturn's orbit, Saturn appears to play a non-negligible 
      role in destabilizing the object's motion. Even if close encounters with Saturn and Neptune are not possible, this object 
      (together with Uranus) currently moves in near resonance with the other three giant planets: 1:7 with Jupiter, 7:20 with 
      Saturn, and 1:2 with Neptune. In order to study the role of these three giant planets on the resonant evolution of (83982) 
      Crantor, we have repeated the calculations considering negligible masses for Jupiter, Saturn and Neptune. This alteration
      removes the associated near resonance. The "toy model" with no Neptune gives very similar results for the evolution of this 
      minor body. Therefore, the role of the near resonance with Neptune can be regarded as negligible. In sharp contrast, the 
      cases of the toy models with no Jupiter or no Saturn are dramatically different. The lack of Jupiter has immediate effects 
      on the results because it amplifies the dynamical effect of Saturn but the absence of Saturn has major effects on the 
      overall orbital evolution of (83982) Crantor. The precession rate decreases by 30\% and the object remains as a complex 
      Uranus horseshoe librator for several 100 kyr. Though (83982) Crantor moves primarily under the influence of the Sun and 
      Uranus, Saturn (mainly), Jupiter and Neptune play an important role by influencing, through torque-induced precession, the 
      position of the asteroid's nodes. Variations in the nodal distance strongly affect the interaction of (83982) Crantor with 
      Uranus and may change or terminate the horseshoe orbit currently observed. This precession of the nodes is the mechanism by 
      which minor planets are placed or removed from horseshoe orbits (Wiegert et al. 1998). On the other hand, the repetitive
      episodes described above (see also Fig. \ref{all}, panel B) in which (83982) Crantor's relative mean longitude librates for 
      several cycles, then circulates for a few more cycles before restarting libration once again are characteristic of a type of 
      dynamical behaviour known as resonance angle nodding (Ketchum et al. 2013). These authors conclude that nodding often occurs 
      when a small body is in an external (near) mean motion resonance with a larger planet. This type of complicated dynamics has 
      been observed in other horseshoe librators. 

      Gallardo (2006) suggested that two objects were trapped in a 1:1 mean motion resonance with Uranus, both moving on horseshoe 
      orbits: (83982) Crantor and 2000 SN$_{331}$. We have just confirmed that (83982) Crantor actually moves in a relatively 
      short-lived horseshoe orbit associated to Uranus. However, we cannot do the same with the other object, 2000 SN$_{331}$. Our
      calculations indicate that this object is not co-orbital but may move in a temporary 10:9 inner resonance with Uranus, i.e., 
      the object completes 10 revolutions around the Sun while Uranus goes around the Sun just 9 times. This result is robust, in 
      principle, but the reliability of the orbit of this object is extremely poor as it is based on just 6 observations with a 
      data-arc span of only 1 day. Therefore, we may say that it is a candidate to move in resonance with Uranus but not a
      co-orbital.

      Given the significant destabilizing role played by Saturn on the orbital evolution of (83982) Crantor, the question 
      concerning the existence of long-term Uranus co-orbitals moving on stable orbits takes a new twist, are they at all possible
      in the light of our present results? In the following section, we present early results on an object that, because of 
      moving in a low eccentricity orbit, may be able to survive as Uranus co-orbital for a longer period of time. In any case, it 
      must be pointed out that longer calculations (several Myr long) suggest that (83982) Crantor may become (600 kyr from now) a 
      long-term Uranus' L$_5$ Trojan. However and given the chaotic nature of the orbit of this object, the reader should take 
      this result with caution.

   \section{2010 EU$_{65}$: a promising Uranus horseshoe candidate}
\begin{figure*}
        \centering
        \includemedia[
          label=Candidate,
          width=\textwidth,height=0.65\textwidth,
          activate=onclick,
          addresource=2010EU65w.mp4,
          flashvars={
            source=2010EU65w.mp4
           &autoPlay=true
           &loop=true
           &controlBarMode=floating
           &controlBarAutoHide=false
           &scaleMode=letterbox
          }
        ]{\includegraphics{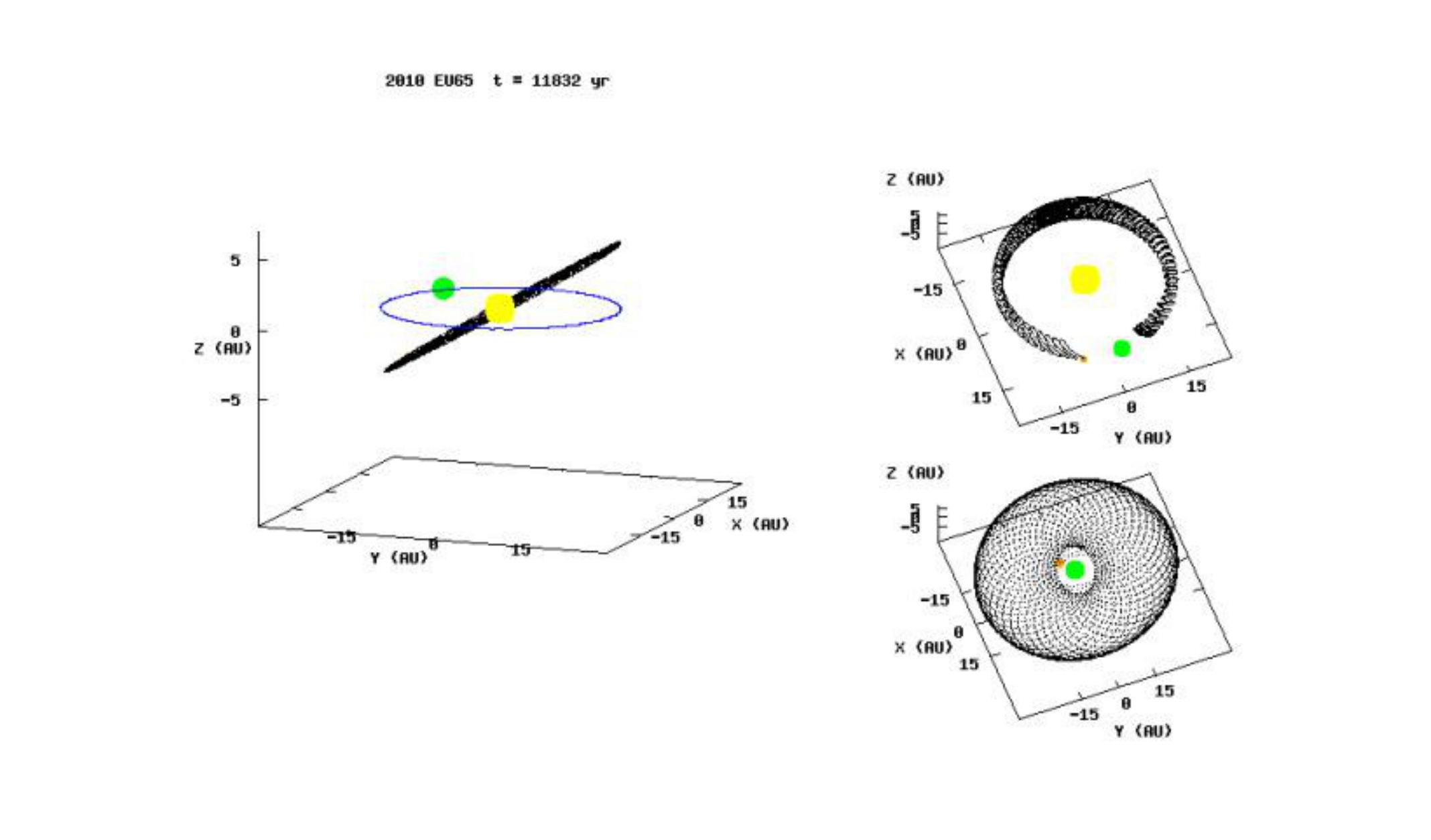}}{VPlayer.swf}
        \PushButton[
           onclick={
             annotRM['Candidate'].activated=true;
             annotRM['Candidate'].callAS('playPause');
           }
        ]{\fbox{Play/Pause}}
        \caption{Three-dimensional evolution of the orbit of 2010 EU$_{65}$ in three different frames of reference: heliocentric 
                 (left), frame co-rotating with Uranus but centered on the Sun (top-right), and Uranocentric (bottom-right). The 
                 orange point represents 2010 EU$_{65}$, the green one is Uranus, and the yellow one is the Sun. The osculating 
                 orbits are outlined and the viewing angle changes slowly to make easier for the reader to visualize the orbital 
                 evolution.} 
        \label{candidate}
     \end{figure*}

      Asteroid 2010 EU$_{65}$ was discovered on March 13, 2010 in images obtained by the European Southern Observatory (ESO) at La
      Silla (Rabinowitz et al. 2012). The object was reobserved in the following days from the same location and also from Cerro 
      Tololo Observatory at La Serena (Rabinowitz et al. 2010). In total, 26 observations with a data-arc span of 85 days. At the 
      time of discovery its apparent magnitude in $R$ was estimated to be 21.2. The Heliocentric Keplerian orbital elements of 
      2010 EU$_{65}$ appear in Table \ref{elementscan}. As a relatively recent discovery, its orbit is poorly constrained and it 
      is included here mainly to encourage follow-up observations. Little is known of the physical properties of this object with 
      the exception of its absolute magnitude of 9.1. This probably suggests a medium-sized object with an estimated diameter in 
      the range 28-90 km, for an albedo range of 0.5-0.05. Its period, 84.16 yr, matches well that of Uranus, 84.32 yr, so it 
      appears to follow a 1:1 resonant orbit with Uranus yet it is classified as a Centaur by the MPC. In general and with the 
      exception of the inclination, its orbit is remarkably similar to that of Uranus. Our calculations for the nominal orbit in 
      Table \ref{elementscan} indicate that 2010 EU$_{65}$ also follows a horseshoe orbit, this time very regular, associated to 
      Uranus. In this case and due to its small eccentricity, the orbit is far more stable than that of 83982 Crantor (2002 
      GO$_{9}$); now the effect of Saturn is rather negligible and the object always remains at a safe distance from Uranus. The 
      immediate future orbital evolution of 2010 EU$_{65}$ is illustrated by the animation displayed in Fig. \ref{candidate} 
      (available on the electronic edition as a high resolution animation or embedded at lower resolution in the pdf file 
      associated to this paper). As in the case of (83982) Crantor, the orbit is presented in three frames of reference: 
      heliocentric (left), co-rotating with Uranus (top-right) and Uranocentric (bottom-right). Our numerical integrations suggest 
      that, in sharp contrast with (83982) Crantor, this object remains in co-orbital motion with Uranus for Myr timescales. Due 
      to its poorly known orbit, we must insist that the object is a mere horseshoe librator candidate (and in dire need of 
      follow-up observations) although all the studied control orbits (with errors below 1\%) give consistent results. 

%
%
         \begin{table}
          \fontsize{8}{11pt}\selectfont
          \tabcolsep 0.35truecm
          \caption{Heliocentric Keplerian orbital elements of 2010 EU$_{65}$ used in this research. The orbit is based on 26 
                   observations spanning a data-arc of 85 days, from 2010-03-13 to 2010-06-06.
                   (Epoch = JD2455300.5, 2010-Apr-14.0; J2000.0 ecliptic and equinox. Source: JPL Small-Body Database.)
                  }
          \begin{tabular}{ccc}
           \hline
            semi-major axis, $a$                        & = & 19.2041 AU \\
            eccentricity, $e$                           & = & 0.05402 \\
            inclination, $i$                            & = & 14.8382 $^{\circ}$ \\
            longitude of the ascending node, $\Omega$   & = & 4.61965 $^{\circ}$ \\
            argument of perihelion, $\omega$            & = & 180.824 $^{\circ}$ \\
            mean anomaly, $M$                           & = & 0.37496 $^{\circ}$ \\
            perihelion, $q$                             & = & 18.1668 AU \\
            aphelion, $Q$                               & = & 20.2414 AU \\
            absolute magnitude, $H$                     & = & 9.1 mag \\
           \hline
          \end{tabular}
          \label{elementscan}
         \end{table}
%
%

%
%
%
%
   \section{Conclusions}
      In this paper we have analyzed the orbital behaviour of Uranus' current horseshoe librator 83982 Crantor (2002 GO$_{9}$)
      numerically in order to better understand its current dynamical status, past dynamics and future evolution as well as to
      gain some insight about its stability. We have also shown that the evolution of this object is mainly controlled by the Sun 
      and Uranus. In fact, close encounters with Uranus generate instability in its orbit and throw the object in and out of the 
      horseshoe dynamical state.

      (83982) Crantor is remarkable in several respects: it is the first known minor body to be trapped in a 1:1 mean motion 
      resonance with Uranus; it currently moves in a complex, horseshoe-like orbit when viewed in a frame of reference co-rotating 
      with Uranus; and it could be the "Rosetta Stone" for understanding why the overall number of Uranus co-orbitals appears to 
      be significantly below that of Jupiter or Neptune. The object is placed and removed from its horseshoe orbit by the 
      mechanism of the precession of the nodes. This precession is accelerated by the perturbative effects of Saturn. The chaotic 
      nature of the orbit of this object constraints the degree of predictability of its dynamical evolution on timescales longer 
      than a few 10 kyr. This strongly suggests that its dynamical age is much shorter than that of the solar system; therefore, 
      (83982) Crantor is unlikely to be a member of a hypothetical primordial population of objects moving in a 1:1 mean motion 
      resonance with Uranus. Horner \& Evans (2006) claimed that Uranus cannot currently efficiently trap objects in the 1:1 
      commensurability even for short periods of time. Our results suggest that, contrary to this view and in spite of the 
      destabilizing role of Saturn, Uranus still can actively capture temporary co-orbitals. Regarding the issue of stability, 
      (83982) Crantor's orbital inclination is close to the edge of one of the stability islands in $i$ identified by Dvorak et 
      al. (2010) but 2010 EU$_{65}$ moves outside the stability islands proposed in that study yet it seems to be more stable than 
      (83982) Crantor. This comparatively better stability strongly suggests that not only inclination but also eccentricity play 
      an important role on the long-term dynamics of these objects. In summary, key questions still remain open and further work 
      is necessary to better understand the complex issue of the stability of Uranian co-orbitals.

   \begin{acknowledgements}
      The authors thank the referee for his/her helpful suggestions regarding the presentation of this paper. The authors would 
      like to thank S.~J. Aarseth for providing the code used in this research. This work was partially supported by the Spanish 
      'Comunidad de Madrid' under grant CAM S2009/ESP-1496 (Din\'amica Estelar y Sistemas Planetarios). We thank M.~J. 
      Fern\'andez-Figueroa, M. Rego Fern\'andez and the Department of Astrophysics of Universidad Complutense de Madrid (UCM) for 
      providing excellent computing facilities. Most of the calculations and part of the data analysis were completed on the 
      'Servidor Central de C\'alculo' of the UCM and we thank S. Cano Als\'ua for his help during that stage. In preparation of 
      this paper, we made use of the NASA Astrophysics Data System and the ASTRO-PH e-print server.
   \end{acknowledgements}

   \bibliographystyle{aa}

\begin{thebibliography}{ }
      \bibitem{AA03} Aarseth, S.~J. 2003,
                     Gravitational N-Body Simulations, 
                     Cambridge University Press, Cambridge, p.\ 27   
      \bibitem{AL07} Alvarez-Candal, A., Barucci, M.~A., Merlin, F., Guilbert, A., \& de Bergh, C. 2007,
                     A\&A, 475, 369 
      \bibitem{AL08} Alvarez-Candal, A., Fornasier, S., Barucci, M.~A., de Bergh, C., \& Merlin, F. 2008,
                     A\&A, 487, 741 
      \bibitem{BA09} Bailey, B.~L., \& Malhotra, R. 2009,
                     Icarus, 203, 155
      \bibitem{BA08} Barkume, K.~M., Brown, M.~E., \& Schaller, E.~L. 2008,
                     AJ, 135, 55
      \bibitem{BR04} Brasser, R., Innanen, K. A., Connors, M., et al. 2004,
                     Icarus, 171, 102
      \bibitem{BR11} Brown, E.~W. 1911,
                     MNRAS, 71, 438
      \bibitem{CH11} Christou, A.~A., \& Asher, D.~J. 2011,
                     MNRAS, 414, 2965
      \bibitem{CO04} Connors, M., Veillet, C., Brasser, R., et al. 2004,
                     Meteoritics Planet. Sci., 39, 1251
      \bibitem{CO05} Connors, M., Stacey, G., Brasser, R., \& Wiegert P. 2005,
                     Plan. Space Sci., 53, 617
      \bibitem{CR07} Cruikshank, D.~P., Barucci, M.~A., Emery, J.~P., Fern\'andez, Y.~R., Grundy, W.~M., 
                     Noll, K.~S., Stansberry, J.~A. 2007,
                     in Protostars \& Planets V, ed. B. Reipurth, D. Jewitt, K. Keil, 
                     University of Arizona Press, Tucson, 879
      \bibitem{DA12} Darwin, G. 1912,
                     MNRAS, 72, 642
      \bibitem{DE09} DeMeo, F.~E., Fornasier, S., Barucci, M.~A., et~al. 2009,
                     A\&A, 493, 283
      \bibitem{D81a} Dermott, S.~F., \& Murray, C.~D. 1981a,
                     Icarus, 48, 1
      \bibitem{D81b} Dermott, S.~F., \& Murray, C.~D. 1981b,
                     Icarus, 48, 12
      \bibitem{DO05} Doressoundiram, A., Barucci, M.~A., Tozzi, G.~P., et~al. 2005,
                     Plan. Space Sci., 53, 1501
      \bibitem{DO07} Doressoundiram, A., Peixinho, N., Moullet, A., Fornasier, S., Barucci, M.~A., 
                     Beuzit, J.-L., \& Veillet, C. 2007,
                     ApJ, 134, 2186 
      \bibitem{DV07} Dvorak, R., Schwarz, R., S\"uli, \'A., \& Kotoulas, T. 2007,
                     MNRAS, 382, 1324
      \bibitem{DV10} Dvorak, R., Bazs\'o, \'A, \& Zhou, L.-Y. 2010,
                     Celest. Mech. Dyn. Astron., 107, 51
      \bibitem{F12a} de la Fuente Marcos, C., \& de la Fuente Marcos, R. 2012a,
                     A\&A, 547, L2
      \bibitem{F12b} de la Fuente Marcos, C., \& de la Fuente Marcos, R. 2012b,
                     A\&A, 545, L9
      \bibitem{F12c} de la Fuente Marcos, C., \& de la Fuente Marcos, R. 2012c,
                     MNRAS, 427, 728
      \bibitem{F12d} de la Fuente Marcos, C., \& de la Fuente Marcos, R. 2012d,
                     MNRAS, 427, L85
      \bibitem{FR12} Fraser, W.~C., \& Brown, M.~E. 2012,
                     ApJ, 749, 33
      \bibitem{GA06} Gallardo, T. 2006,
                     Icarus, 184, 29
      \bibitem{GA77} Garfinkel, B. 1977,
                     AJ, 82, 368
      \bibitem{GI70} Giacaglia, G.~E.~O. 1970,
                     in Periodic Orbits, Stability and Resonances, ed. G.~E.~O. Giacaglia, 
                     Dordrecht, Reidel, 515
      \bibitem{GI02} Gilmore, A.~C., Pravec, P., Helin, E.~F., et~al. 2001,
                     MPEC 2002-H03
      \bibitem{GI96} Giorgini, J.~D., Yeomans, D.~K., Chamberlin, A.~B., et~al. 1996,
                     BAAS, 28, 1158
      \bibitem{HO93} Holman, M.~J., \& Wisdom, J. 1993,
                     AJ, 2015, 1987
      \bibitem{HO06} Horner, J., \& Evans, N. 2006,
                     MNRAS 367, L20
      \bibitem{HO04} Horner, J., Evans, N.~W., \& Bailey, M. E. 2004,
                     MNRAS 354, 798
      \bibitem{KA04} Karlsson, O. 2004,
                     A\&A, 413, 1153
      \bibitem{KE13} Ketchum, J.~A., Adams, F.~C., \& Bloch, A.~M. 2013,
                     ApJ, 762, 71
      \bibitem{MA91} Makino, J. 1991,
                     ApJ, 369, 200
      \bibitem{MA03} Marzari, F., Tricarico, P., \& Scholl, H. 2003,
                     A\&A, 410, 725
      \bibitem{MI96} Michel, P., Froeschl\'e, C., \& Farinella, P. 1996,
                     A\&A, 313, 993 
      \bibitem{MI06} Mikkola, S., Innanen, K., Wiegert, P., Connors, M., \& Brasser, R. 2006,
                     MNRAS, 369, 15
      \bibitem{MI89} Milani, A., Carpino, M., Hahn, G., \& Nobili, A.~M. 1989,
                     Icarus, 78, 212
      \bibitem{MU99} Murray, C.~D., \& Dermott, S.~F. 1999,
                     Solar System Dynamics,
                     Cambridge University Press, Cambridge, p.\ 97
      \bibitem{NA99} Namouni, F. 1999,
                     Icarus, 137, 293
      \bibitem{NE02} Nesvorn\'y, D., \& Dones, L. 2002,
                     Icarus, 160, 271
      \bibitem{OR03} Ortiz, J.~L., Guti\'errez, P.~J., Casanova, V., \& Sota, A. 2003,
                     A\&A, 407, 1149
      \bibitem{RA61} Rabe, E. 1961,
                     AJ, 66, 500
      \bibitem{RA10} Rabinowitz, D., Tourtellotte, S., \& Marsden, B.~G. 2010,
                     MPEC 2010-H80 
      \bibitem{RA12} Rabinowitz, D., Schwamb, M.~E., Hadjiyska, E., Tourtellotte, S. 2012,
                     AJ, 144, 140 
      \bibitem{SM80} Smith, B.~A., Reitsema, H.~J., Fountain, J.~W., \& Larson, S.~M. 1980,
                     BAAS, 12, 727
      \bibitem{ST98} Standish, E.~M. 1998, 
                     JPL Planetary and Lunar Ephemerides, DE405/LE405,
                     Interoffice Memo. 312.F-98-048, Jet Propulsion Laboratory, Pasadena, California
      \bibitem{ST08} Stansberry, J., Grundy, W., Brown, M., Cruikshank, D., Spencer, J., Trilling, D., \& Margot, J.-L. 2008,
                     in The Solar System Beyond Neptune, ed. M.~A. Barucci, H. Boehnhardt, D.~P. Cruikshank, A. Morbidelli,
                     University of Arizona Press, Tucson, 161
      \bibitem{SY81} Synnott, S.~P., Peters, C.~F., Smith, B.~A., \& Morabito, L.~A. 1981,
                     Science, 212, 191
      \bibitem{TE03} Tegler, S.~C., Romanishin, W., \& Consolmagno, G.~J. 2003,
                     ApJ, 599, L49
      \bibitem{TH59} Thuring, B. 1959,
                     AN, 285, 71
      \bibitem{TI02} Ticha, J., Tichy, M., Haver, R., et~al. 2002,
                     MPEC 2002-L55
      \bibitem{WA12} Wajer, P., \& Kr\'olikowska, M. 2012,
                     Acta Astronomica, 62, 113
      \bibitem{WE74} Weissman, P.~R., \& Wetherill, G.~W. 1974,
                     AJ, 79, 404
      \bibitem{WI97} Wiegert, P., Innanen, K. A., \& Mikkola, S. 1997,
                     Nature, 387, 685
      \bibitem{WI98} Wiegert, P.~A., Innanen, K.~A., \& Mikkola, S. 1998,
                     AJ, 115, 2604
      \bibitem{WI00} Wiegert, P., Innanen, K., \& Mikkola, S. 2000,
                     AJ, 119, 1978
   \end{thebibliography}

\end{document}